\documentclass[prd,preprintnumbers,twocolumn,eqsecnum,floatfix,a4paper,nofootinbib,superscriptaddress]{revtex4}
\usepackage{color}
\usepackage{calc}
\usepackage{amsmath,amssymb,graphicx}
\usepackage{amssymb,amsmath}
\usepackage{bm}
\usepackage{microtype}
\usepackage{booktabs}
\usepackage{times}
\usepackage{subfigure}
\usepackage[varg]{txfonts}
\usepackage[utf8]{inputenc}
\definecolor{LinkColor}{rgb}{0.75, 0, 0}
\definecolor{CiteColor}{rgb}{0, 0.5, 0.5}
\definecolor{UrlColor}{rgb}{0, 0, 0.75}
\allowdisplaybreaks
\DeclareFontFamily{OT1}{pzc}{}
\DeclareFontShape{OT1}{pzc}{m}{it}{<-> s * [1.10] pzcmi7t}{}
\DeclareMathAlphabet{\mathpzc}{OT1}{pzc}{m}{it}
\usepackage[breaklinks, plainpages=false, colorlinks=true, anchorcolor=cyan, linkcolor=red, citecolor=blue, urlcolor=magenta, bookmarks=false]{hyperref}
\usepackage{natbib}
\usepackage{blindtext}
\begin{document}
\renewcommand{\thefigure}{\arabic{figure}}
\setcounter{figure}{0}
\def\I{{\rm i}}
\def\E{{\rm e}}
\def\D{{\rm d}}

\title{Modified Gravity Theories \\in Light of the Anomalous Velocity Dispersion of NGC1052-DF2}

\author{Tousif Islam}
\email[]{tousifislam24@gmail.com}
\affiliation{ International Centre for Theoretical Sciences, Tata Institute of Fundamental Research, Bangalore- 560012, India}
\affiliation{ Center for Scientific Computation and Visualization Research (CSCVR), University of Massachusetts (UMass) Dartmouth, Dartmouth, MA-02740, USA}

\author{Koushik Dutta}
\email[]{koushik.physics@gmail.com}
\affiliation{ Theory Divison, Saha Institute of Nuclear Physics, HBNI,1/AF Bidhannagar, Kolkata- 700064, India}

\begin{abstract}
Recent observations of ultra-dwarf galaxy NGC1052-DF2 started an interesting discussion between dark matter hypothesis and modified gravity theories. Reported low velocity dispersion ($ <10.5$ km/s at $90\%$ confidence level) derived from the kinematic data of 10 globular clusters in the galaxy points towards an extraordinarily low dynamical mass ($\sim$ $3.4 \times 10^{8} M_{\odot}$) which is of the same order of the luminous mass ($\sim$ $2.0 \times 10^{8} M_{\odot}$) in the galaxy. This has been interpreted as the first evidence of a galaxy `without Dark Matter'. It has been argued that dark matter is not necessarily coupled to the the baryonic mass on the galactic scale and poses a challenge to modified gravity theories. We explore the dynamics of NGC1052-DF2 within the context of four popular alternative theories of gravity [Modified Newtonian Dynamcies (MOND), Weyl Conformal gravity, Modified gravity (MOG)/Scalar-Tensor-Vector Gravity (STVG) and Verlinde's Emergent gravity] and present the analysis of detailed radial variation of the velocity dispersion. We demonstrate that the dispersion data of NGC1052-DF2 is fully consistent with modified gravity paradigm (as well as with general relativity without dark matter). We reach similar conclusion for the ultra-dwarf NGC1052-DF4 which has been claimed to be the second candidate for galaxies `without Dark Matter'.
\pacs{}
\end{abstract}
\maketitle
\section{\textbf{Introduction}} Recently, van Dokkum \textit{et al.} (vD18a, vD18b) \cite{van2018galaxy,van2018revised} used kinematic data of ten globular clusters (GCs) of the ultra-diffuse galaxy NGC1052-DF2 as bright tracers of its potential and reported a  line-of-sight velocity dispersion of $\sigma \sim 7.8_{-2.2}^{+5.6}$ km/s. They inferred a total dynamical mass of $\sim 3.4 \times 10^{8} M_{\odot}$ (where $M_{\odot}$ is the solar mass) within the radius of 7.6 kpc and a luminous mass of $\sim 2.0 \times 10^{8} M_{\odot}$ within the same radius. This implies a ratio for $M_{dyn}/M_{stars}$ of order unity. As this ratio is typically of the order $\sim 10^{2}-10^{3}$ \cite{behroozi2013average}, the authors concluded that the galaxy is consistent with having little to no dark matter. They further postulated that the effect of Dark matter need not to be coupled to the baryonic mass of the galaxies, thus challenging the notion of alternative theories of gravity and the phenomenologically established Radial-Acceleration-Relation (RAR) between Newtonian baryonic acceleration and the observed acceleration \cite{McGaugh} for galaxies. 

Several studies, however, have raised questions regarding the accuracy of estimated intrinsic velocity dispersion value and the statistical methods employed; mainly due to the paucity of kinematic sample. Emsellem \textit{et. al} separately derived a velocity dispersion of $16.3\pm5$ km/s for the stellar components of NGC1052-DF2 using the MUSE@VLT spectograph \cite{emsellem2019ultra}. Another independent analysis by Martin \textit{et. al} \cite{martin2018current} using the same data of vD18a reports a mean observed bi-weighted dispersion value of $14.3\pm3.5$ km/s. Both these estimates are significantly higher than the value quoted by vD18a. However, a different study by Danieli \textit{et. al} \cite{danieli2019still} have found a stellar velocity dispersion of $\sigma \sim 8.4\pm2.1$ km/s, consistent with the values derived from the GCs in vD18a. More recently Haghi \textit{et al.}\cite{haghi2019new} have re-analyzed the velocity data of vD18a and obtained an estimate of $2.4-18.8$ km/s for the intrinsic velocity dispersion of NGC1052-DF2 at $2\sigma$ confidence level. All of these studies have maintained the view that the current data do not allow to strongly claim whether the galaxy is devoid of dark matter or not.

Another major issue is the distance to NGC1052-DF2 from us. vD18a had obtained a distance ($D$) of $20$ Mpc (vD18b had revised the value to $19\pm1.7$ Mpc). However, this estimate remains highly disputed as Trujillo \textit{et. al} \cite{trujillo2019distance} reports a different estimation of  $D\sim13.2$ Mpc which brings the estimated baryonic mass around $\sim 6\times 10^{7}$ $M_{\odot}$ and the ratio $M_{dyn}/M_{stars}$ becomes high enough such that the galaxy can be treated as `normal'. 

Even if one ignores the dispute regarding the distance estimates and statistical uncertainties and assumes the galaxy to have no dark matter, there still lies a number of puzzles that 
remain difficult to explain in the typical $\Lambda$CDM (general relativity along with dark matter and dark energy) model. The standard cosmology and stellar-to-halo mass relation (SHMR) are found to be in tension with the observed baryonic and dynamical mass estimates in DF2 \cite{haslbauer2019ultra,wasserman2018deficit}. Furthermore, the astrophysical formation channel for such `dark matter deficit' galaxies is not clear to date \cite{nusser2019scenario,ogiya2018tidal,leigh2019collisional,silk2019ultra}.

In this paper, we therefore set out to revisit the question whether this low velocity dispersion estimates and the `apparent' lack of dark matter is consistent (or at odds) with the modified gravity theories. We choose four typical alternative theories of gravity: namely, Modified Newtonian gravity (MOND) \cite{mond1,famaey2012modified}, 
Scalar-Tensor-Vector Gravity (STVG) or MOdified Gravity (MOG) \cite{mog}, Weyl Conformal gravity \cite{weyl1,weylrot5} and Verlinde's Emergent Gravity \cite{verlinde2017emergent}. All these modified theories of gravity attempt to explain the dynamics of galaxies and globular clusters without invoking the `exotic' dark matter. Modified gravity theories, except Emergent gravity, have remained extremely successful to explain the observed rotation curves of a large selection of galaxies (\cite{sanders2002modified,gentile2011things} for MOND; \cite{weylrot1,weylrot2,weylrot3,weylrot4,kt2018} for Weyl gravity; \cite{moffat2013mog,moffat2015rotational,moffat2011testing} for MOG), dispersion profiles of galactic globular clusters (\cite{scarpa2003using,scarpa2004using,scarpa2007using} for MOND; \cite{islam2018globular} for Weyl gravity; \cite{moffattoth} for MOG) and the phenomenological RAR for galaxies (\cite{ghari2019radial,kt2018} for MOND; \cite{o2019radial} for Weyl gravity; \cite{green2019modified} for MOG).  Emergent gravity, on the other hand, enjoys success in explaining the rotation curves of dwarf galaxies \cite{diez2018verlinde}  but finds itself inconsistent with RAR \cite{lelli2017testing}.

Famaey \textit{et al.} \cite{famaey2018mond}, Kroupa \textit{et al.} \cite{kroupa2018does} and Haghi \textit{et al.} \cite{haghi2019new} have recently explored the dynamics of NGC1052-DF2 in the MOND paradigm and concluded that the current data are insufficient to rule out MOND scenarios. Moffat and Toth \cite{moffat2018ngc} found MOG to be consistent with the quoted overall dispersion value in vD18a. However, these studies lack a detailed comparison between the observed dispersion velocities for individual GCs and the predicted values, which we will investigate thoroughly. 

The outline of the remaining paper is as follows. We begin with the baryonic mass model of NGC1052-DF2 (Section \ref{sec2}), which is followed by a brief introduction of MOND, Weyl Conformal gravity, MOG and Emergent Gravity theory (Section \ref{sec3}). We then present the formulation of velocity dispersions in a spherically symmetric system like NGC1052-DF2 (Section \ref{sec4}) and proceed to compare the predicted dispersion profiles in each modified theories of gravity with the observed one (Section \ref{sec5}). We further discuss implications of the results, point out possible caveats of the analysis and conclude (Section \ref{sec6}).
\section{\textbf{Mass Profile of NGC1052-DF2}} 
\label{sec2}
Van Dokkum \textit{et. al} (vD18a) \cite{van2018galaxy} have modelled the surface brightness of the ultra-diffuse galaxy NGC1052-DF2 using a two-dimensional S\'{e}rsic profile, with index $n = 0.6$, effective radius $R_e = 2.2$ kpc and total luminosity $L=1.2\times10^{8}$ $L_{\odot}$.  Cohen \textit{et. al}  \cite{cohen2018dragonfly} also have fitted the surface brightness profile of  NGC1052-DF2 using a slightly different Sersic model characterized by $n=0.55$ and $R_e=1.8$ kpc. Both the fits assume a distance to the galaxy $D=20$ Mpc. These parameter values indicate a total mass of $M\sim2.0\times 10^{8}M_{\odot}$ within a radius $7.6$ kpc (assuming standard stellar population modelling). Moffat and Toth \cite{moffat2018ngc} have found that the resultant mass density can be closely approximated as: 
\begin{equation}
\rho_{ser}=\frac{40 \varSigma_0}{63 R_e} exp\Big[   -\left(\frac{11r}{10R_e}\right)^{4/3} \Big],
\label{mass}
\end{equation}
where $\varSigma_0=1.25 \times 10^7 M_{\odot}/kpc^2$ is the characteristic surface mass density and $R_e=2.0$ $kpc$ is the effective radius.
\begin{figure}[ht]
	\centering
	\includegraphics[width=0.95\linewidth]{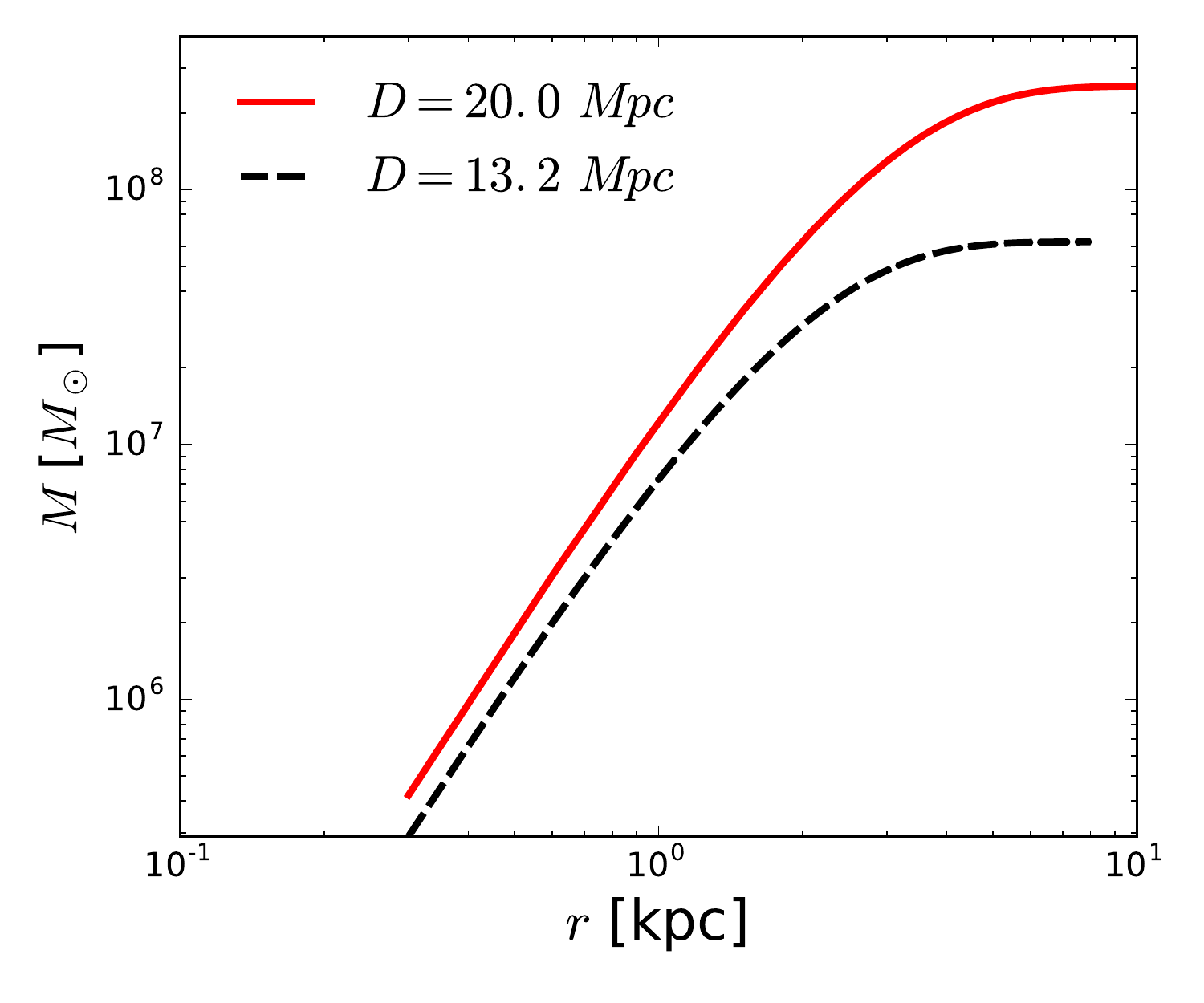}
	\caption[]{\textbf{Enclosed baryonic (stellar) mass profile of NGC1052-DF2}: Solid red line denotes the enclosed mass assuming a distance to galaxy $D=20$ Mpc. However, the estimated mass decreases if the galaxy is much closer. In dashed black line, we plot the mass profile for $D=13.2$ Mpc.}
	\label{fig1}
\end{figure}
However, an independent distance measurement of NGC1052-DF2 by Trujillo \textit{et. al} \cite{trujillo2019distance} have reported a distance of  $13.2$ Mpc. This particular estimate yields an estimated total mass around $\sim 6\times 10^{7}$ $M_{\odot}$ and effective radius $R_e=1.4 \pm 0.1$ kpc. In Figure \ref{fig1}, we show the baryonic stellar mass profile for two different distance measurements.

\section{Modified Gravity Theories}
\label{sec3}
\paragraph{\textbf{Modified Newtonian Dynamcies (MOND)}:---}In Modified Newtonian Dynamcies (MOND) \cite{mond1,famaey2012modified} scenarios, the Newtonian acceleration is modified through an interpolating function $\mu$ such that
\begin{equation}
\mu \left(\frac{a}{a_0}\right) a = a_{N}.
\end{equation}
The interpolating function $\mu(x) \approx x$ when $x \ll 1$ and $\mu(x) \approx 1$ when $x \gg 1$. Therefore, when the gravitational acceleration is high, Newtonian behavior is recovered. The quantity $a_0$ denotes a critical value below which Newtonian gravity breaks down. One can use different functional forms of the interpolating function $\mu(x=\frac{a}{a_0})$. In this paper, we stick to the following `standard' form: $\mu (x) = \frac{x}{\sqrt{(1 + x^{2})}}$, with $a_0$ = $1.21 \times 10^{-10} m/s^{2}$. The MOND acceleration then becomes \cite{mond1}
\begin{equation}
a_{MOND} =  \frac{a_{N}}{\sqrt{2}}\left( 1 +  \Big( 1 + \left(2a_0/a_{N}\right)^2 \Big)^{1/2} \right)^{1/2},
\end{equation}
where $a_{N}=\frac{GM(r)}{r^{2}}$ is the Newtonian acceleration associated with the baryonic mass.\\
\paragraph{\textbf{Weyl Conformal Gravity}:---} In addition to the general coordinate invariance and equivalence principle, Weyl conformal gravity \cite{weyl1,weylrot5} employs the principle of local conformal invariance of the space-time in which the action remains invariant under the transformation $g_{\mu \nu} (x) \rightarrow \Omega^{2}(x) g_{\mu \nu} (x)$, where  $g_{\mu \nu}$ is the metric tensor and $\Omega(x)$ is a smooth strictly positive function. This leads to a unique action $I_{w}=-\alpha_{g} \int d^{4}x \sqrt{-g} C_{\lambda\mu\nu\kappa} C^{\lambda\mu\nu\kappa}$ with $\alpha_{g}$ is a dimensionless coupling constant and $C_{\lambda\mu\nu\kappa}$ is the Weyl tensor \cite{weyl1918}. The action gives rise to a fourth order field equation \cite{weylrot5}. For a spherically symmetric mass distribution, one can show that the final expression of the resultant acceleration will read \cite{weylcluster2,weylcluster1}:
\begin{align}
&
\begin{aligned}[t]
&a(r)  \\
&= G \left[-{I_0(r)\over r^2} + {1\over R_0^2}\left({I_2(r)\over 3 r^2} - {2\over 3} r E_{-1}(r) 
- I_0(r)\right) \right] \\
& + {GM_0\over R_0^2}  - \kappa c^2 r \; .
\end{aligned}
\label{weylequ}
\end{align}
where $I_n(r)=4\pi \int_0^r \rho(x)x^{n+2} dx$ and $E_n(r)=4\pi \int_r^{+\infty} \rho(x)x^{n+2}dx$ are the interior and exterior mass moments respectively. Previous fits to galaxy rotation curves \citep{weylrot2,weylrot1,weylrot3,weylrot4} yields the following values for the Weyl gravity parameters: $R_0=24$~kpc and $M_0= 5.6\times 10^{10}M_\odot$ and $\kappa = 9.54 \times 10^{-54} $ $cm^{-2}$.\\
\paragraph{\textbf{Modified gravity (MOG) theory}:---}Scalar-Tensor-Vector Gravity (STVG), otherwise known as modified gravitational (MOG) theory, includes a massive vector field $\phi_{\mu}$ and three scalar fields $G$, $\mu$ and $\omega$. While  $G$ denotes the dynamical version of the Newtonian gravitational constant,  $\phi_{\mu}$ and $\omega$ represents the  mass  of  the  vector  field \cite{mog}. The acceleration in a spherically symmetric mass distribution in MOG takes the following form:
\begin{align}
\begin{aligned}
	a_{MOG}=&- \int_{0}^{r}    dr'\frac{2 \pi Gr'}{\mu r^2}\rho(r')\Big\{2(1+\alpha) \\
	& + \alpha(1+\mu r)[e^{-\mu(r+r')}-e^{-\mu(r-r')}] \Big\}\\
	& -\int_{r}^{\infty}    dr'\frac{2 \pi Gr'}{\mu r^2}\rho(r')\alpha\\ 
	&\times \Big\{ (1+\mu r)[e^{-\mu(r+r')}-(1-\mu r)e^{-\mu(r'-r)}]\Big\}
\end{aligned}
\end{align}
where $\alpha$ and $\mu$ controls the strength and range of the attractive force, and are generally functions of the mass enclosed in a system \cite{moffat2009fundamental}; $G$ is the Newtonian gravitational constant. For a system like NGC 1052-DF2, the corresponding values would be: $\alpha=1.30$ and $\mu=0.443$ $kpc^{-1}$ \cite{moffat2018ngc}.\\
\paragraph{\textbf{Emergent Gravity theory}:---} Emergent gravity or Entropic gravity postulates that, unlike other forces (weak, strong and electromagnetic) in nature, gravity emerges from an underlying microscopic theory connecting thermodynamics and quantum information theory \cite{verlinde2017emergent}. For a mass distribution exhibiting spherical symmetry, the acceleration profile in Emergent gravity is given by:
\begin{equation}
a_{EG} = \bigtriangledown\phi_{N} + \Big[(\bigtriangledown \phi_{N}) a_{v} \Big]^{1/2},
\end{equation}
where $a_{N}=\bigtriangledown \phi_{N}$ is the Newtonian gravity acceleration (from baryonic mass) and $a_{v}$ is defined as:
\begin{equation}
a_{v} =  \frac{a_0}{M(r)}\frac{d(M(r)r)}{dr},
\end{equation}
where $a_0$ is analogous to the MOND acceleration scale. Here, we take $a_0$ = $1.21 \times 10^{-10} m/s^{2}$ (the same value used for the $a_0$ in MOND). However, we would like to point out that Verlinde \cite{verlinde2017emergent} used a slightly different value for $a_0$. However, this difference would not change our result.
\section{\textbf{Dispersion Profile of Spherically Symmetric Objects}}
\label{sec4}
For non-rotating systems with spherically symmetric mass distributions, the velocity dispersion can be computed by solving the Jeans equation \cite{bt1987}
\begin{equation}
\frac{\partial(\rho(r)\sigma^2(r))}{\partial r} +\frac{2 \rho(r) \xi \sigma^{2}(r)}{r} = \rho(r) a(r),
\label{eqn15}
\end{equation}
where $r$ is the radial distance from the center of the object and $\rho(r)$ is the radial density function. We assume $\lim\limits_{r\rightarrow\infty}\rho(r)\sigma^2(r)=0$ and anisotropy parameter $\xi=0$. It allows us to write
\begin{equation}
\sigma^2(r)=\frac{1}{\rho(r)}\int\limits_r^\infty\rho(r')a(r')~dr'.
\label{eqn16}
\end{equation}
The corresponding  projected line-of-sight (LOS) velocity dispersion is then straightforwardly written:
\begin{equation}
\sigma_\mathrm{LOS}^2(R)=\frac{\int_R^\infty r\sigma^2(r)\rho(r)/\sqrt{r^2-R^2}~dr}{\int_R^\infty r\rho(r)/\sqrt{r^2-R^2}~dr},
\label{eqn17}
\end{equation}
where $R$ is the projected distance from the center of the object.\\
\begin{table}[h]
	\centering
	\caption[Milky Way mass model]{\textbf{Reduced chi-square values as goodness-of-fits for different theories of gravity}. $D=20$ $Mpc$ and no dark matter are assumed.}
	\label{T1}
	\begin{tabular}{c c}
		\hline 
		\hline
		&$\chi^{2}/dof$ \\
		General Relativity (GR) without dark matter &1.82\\
		Modified Gravitational Theory (MOG) & 1.47\\
		Modified Newtonian Dynamcies (MOND) & 3.60\\
		MOND with EFE & 2.01\\
		Weyl Conformal Gravity & 5.00\\
		Emergent Gravity & 9.99\\
		\hline 
		\hline
	\end{tabular} 
\end{table}

\section{\textbf{Results}}
\label{sec5}
\begin{figure}[h!]
	\centering
	\includegraphics[width=1.0\linewidth]{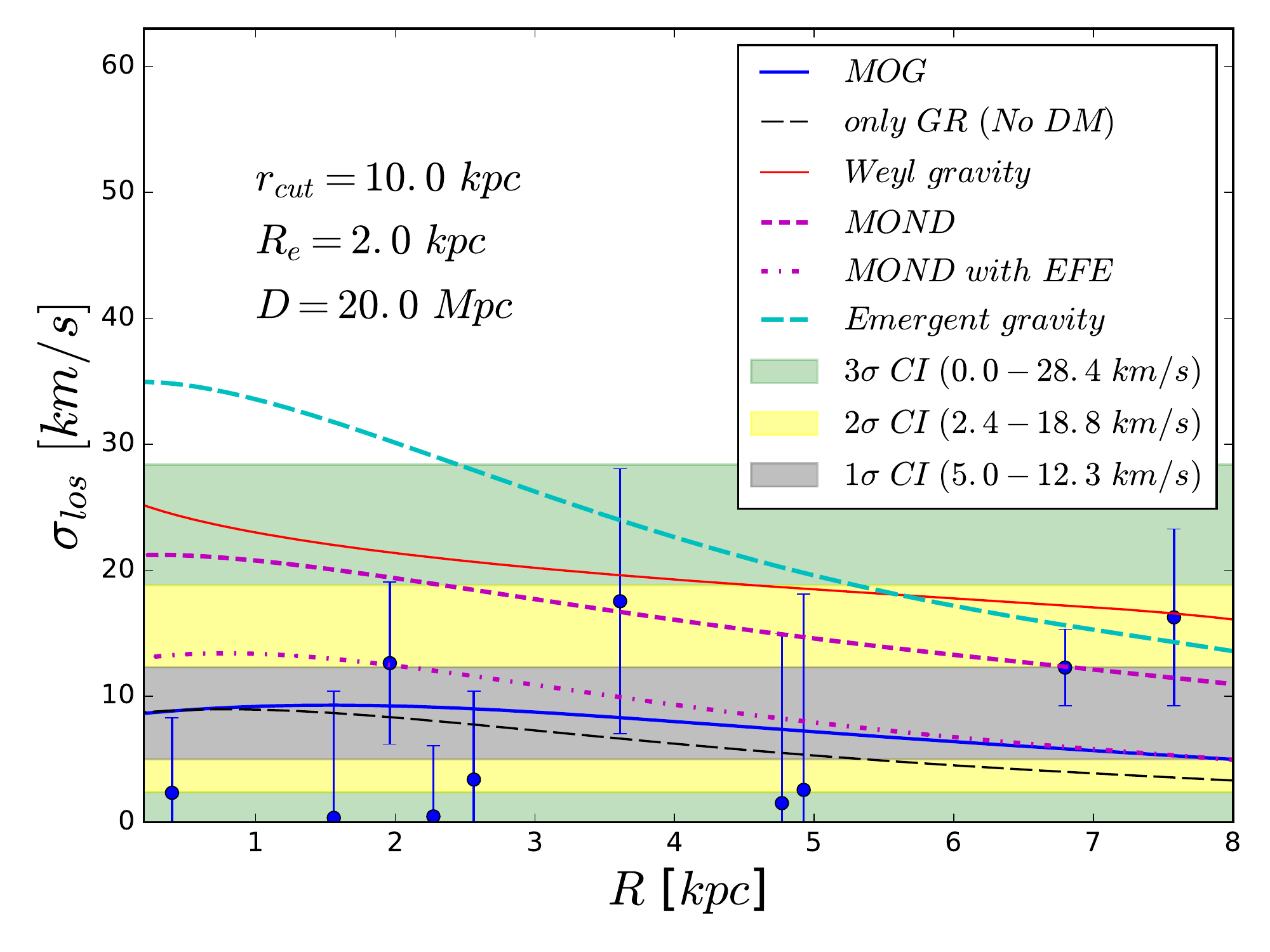}
	\caption[]{\textbf{Line-of-sight velocity dispersion profile of NGC1052-DF2 for $D=20$ Mpc}: The blue dots with error-bars denote the individual dispersion measurements. $1\sigma$ (gray), $2\sigma$ (yellow) and $3\sigma$ (green) confidence limits  \cite{haghi2019new} on the observed velocity dispersion data are colored. On top of these, we plot the predicted Weyl gravity dispersion profile in red solid line, MOND profile in short dashed magenta line, MOND with EFE in dash-dotted magenta line, MOG profile in thick solid blue and Emergent gravity in thick cyan long dashed line. For the GR profile, we assume no dark matter and plot it as a black long dashed line.}
	\label{fig2}
\end{figure}
If NGC1052-DF2 is located at a distance of $D\sim20$ MPc, it is likely to be associated with a much larger galaxy NGC 1052 ($ M\sim10^{11}M_{\odot}$), separated by a distance of only 80 kpc. Due to the gravitational pull of NGC1052, the outer region of the ultra-diffuse galaxy NGC1052-DF2 might have been stripped of matter; resulting a trimmed mass profile. We generically model this scenario by performing a sharp trimming at a certain radius $r_{cut}$. The mass density (presented in Equation \ref{mass}) beyond this radius is set to be zero. In this work, we adopt a phenomenological value of $r_{cut}=10$ kpc (keeping in mind that the radial distance of the last observed GC is $\sim$ 8 kpc).

Such simplistic approach might not be sufficient to capture the effects of the external gravitational fields on the dynamics of NGC 1052-DF2 in MOND. We therefore incorporate the External Field Effect (EFE) unique to MOND \cite{famaey2018mond,haghi2019new}, in addition to the isolated MOND case (i.e. ignoring EFE completely), in our computation. The internal acceleration of NGC 1052-DF2 is $\sim$0.12$a_0$ whereas the external acceleration due to the host NGC 1052 is $\sim$0.15$a_0$ \cite{famaey2018mond}. We, however, assume the extreme case where the internal acceleration is dominated by the external field. We set $a_{ext}=0.5$$a_{0}$. The interpolating function then becomes:
\begin{equation}
\mu(\frac{a}{a_0})\rightarrow \mu(\frac{a+a_{ext}}{a_0})=\frac{a_{ext}}{\sqrt{a_{ext}^2+a_0^2}}.
\end{equation}
The true MOND dispersion profile thus lies somewhere in between the predicted profiles of the isolated MOND and MOND with extreme EFE.

We first present a comparison between the data \cite{wasserman2018deficit} and the predicted line-of-sight (los) velocity dispersion profiles in modified gravity theories in Figure \ref{fig2}. The individual GC velocity dispersion measurements are shown in blue circles along with their quoted error-bars. On top of that, we highlight the $1\sigma$, $2\sigma$ and $3\sigma$ confidence intervals on the observed velocity dispersion as inferred in the latest study by Haghi \textit{et al.} \cite{haghi2019new} employing full-fledged Monte-Carlo method with original radial velocity data from vD18 \cite{van2018galaxy,van2018revised}. The predicted modified gravity dispersion profiles from the inferred baryonic mass (long dashed black line for GR; solid red line for Weyl gravity; short dashed magenta line for MOND; dash-dotted magenta line for MOND with EFE; solid thick blue line for MOG; and thick cyan long dashed line for Emergent gravity) are then superimposed. We report the reduced chi-square values ($\chi^{2}/dof$ where $dof$ is the degree of freedom) for different theories of gravity as a goodness of fits between prediction and observation in Table \ref{T1}. 

It is clear that the predicted profiles of four gravity models (GR/MOND/MOND with EFE/MOG) lie within the $2\sigma$ confidence interval of the observed data. While $\sim$ 6 and $\sim$ 4 of the observed velocities lie on the MOND and MOND with EFE  profile respectively, both GR (without DM) and MOG manages to cross $\sim$ 7 of the observed velocities. However, Weyl gravity ($\chi^{2}/dof=5.00$) (that matches the data only at 3$\sigma$ confidence level) and MOND ($\chi^{2}/dof=3.60$) seems to be a bit off from the data, with a spatially averaged $\sigma \sim 14-22$ km/s, compared to an observed value of $\sigma \sim 5-10$ km/s whereas both GR and MOG appears more close to the data with reduced chi-square value $\chi^{2}/dof=1.82$ and $\chi^{2}/dof=1.47$ respectively. Weyl gravity manages to touch only 3 of the observed velocity data -points. We further notice that including EFE in MOND results in significant reduction in dispersion values. However, our computation reflects an extreme case of EFE. The true MOND profile for NGC 1052-DF2, under the external field of $a_{ext}=0.15a_0$, will result a reduced chi-square value in between $2.01$ (for extreme EFE) and $3.60$ (for isolated MOND).

\begin{figure}[ht]
	\centering
	\includegraphics[width=1.0\linewidth]{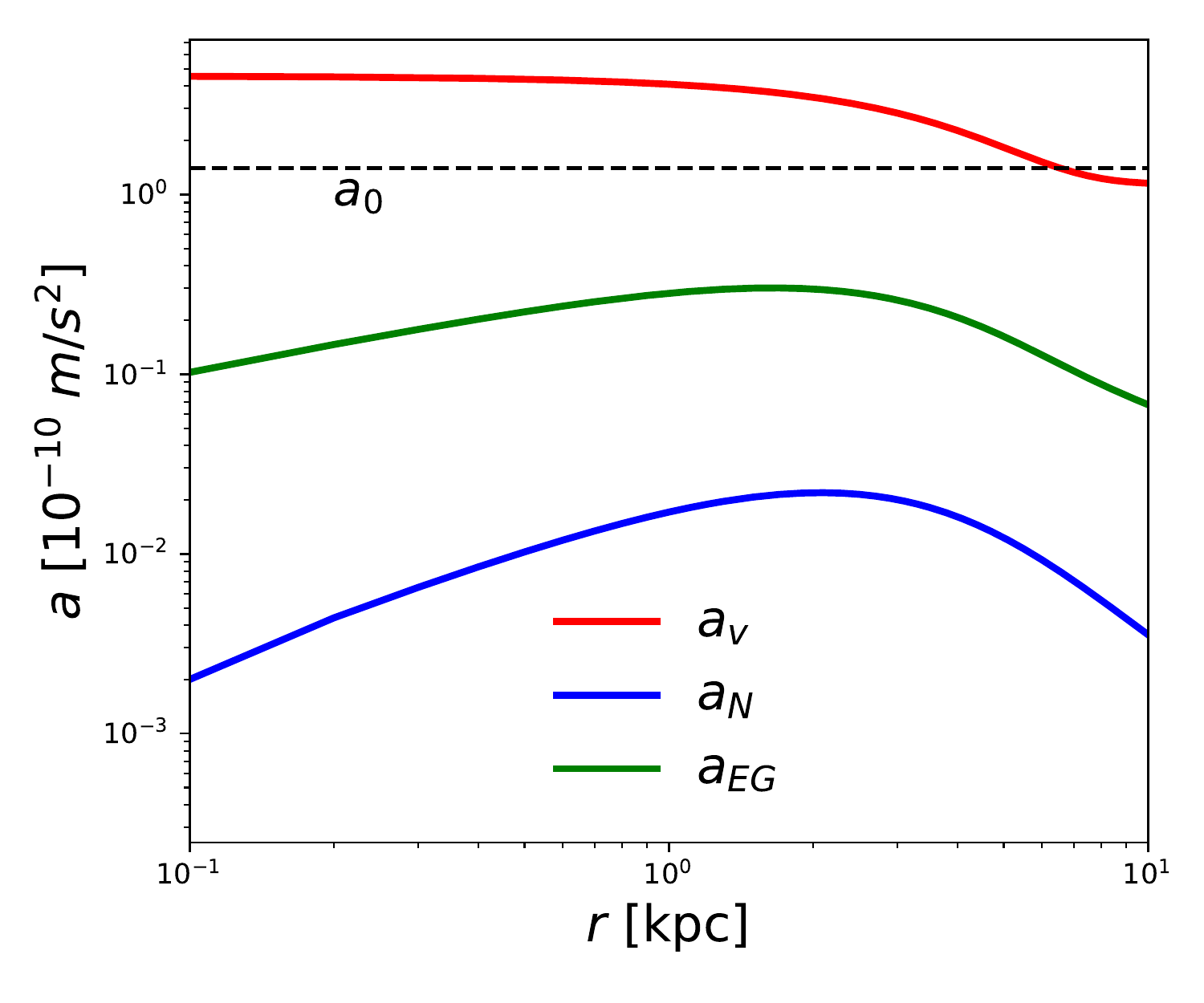}
	\caption[]{\textbf{Radial variation of the additional acceleration component $a_v$ in Emergent gravity for NGC1052-DF2}. We identify that its relatively higher value ($a_v$; red) than the Newtonain acceleration ($a_N$; blue) is the reason why Emergent gravity ($a_{EG}$; green) gives a relatively poor match with data.}
		\label{fig3b}
	\end{figure}

On the other hand, Emergent gravity cannot even match the data at $3\sigma$ confidence interval and fails to account for the most of the points in the interior of the galaxy and yields a chi square value $\chi^{2}/dof=9.99$ way higher than other gravity theories. The theory predicts a spatially averaged $\sigma \sim 20-26$ km/s. We realize that the additional acceleration component in the Emergent gravity $a_{v} =  \frac{a_0}{M(r)}\frac{d(M(r)r)}{dr}$ takes relatively higher values (almost two-three orders of magnitude higher) than its Newtonian counterpart ($a_N$) (Figure \ref{fig3b}). This forces the predicted Emergent gravity profile to assume much higher dispersion values than in data.

\begin{figure}[h]
	\centering
	\includegraphics[width=0.8\linewidth]{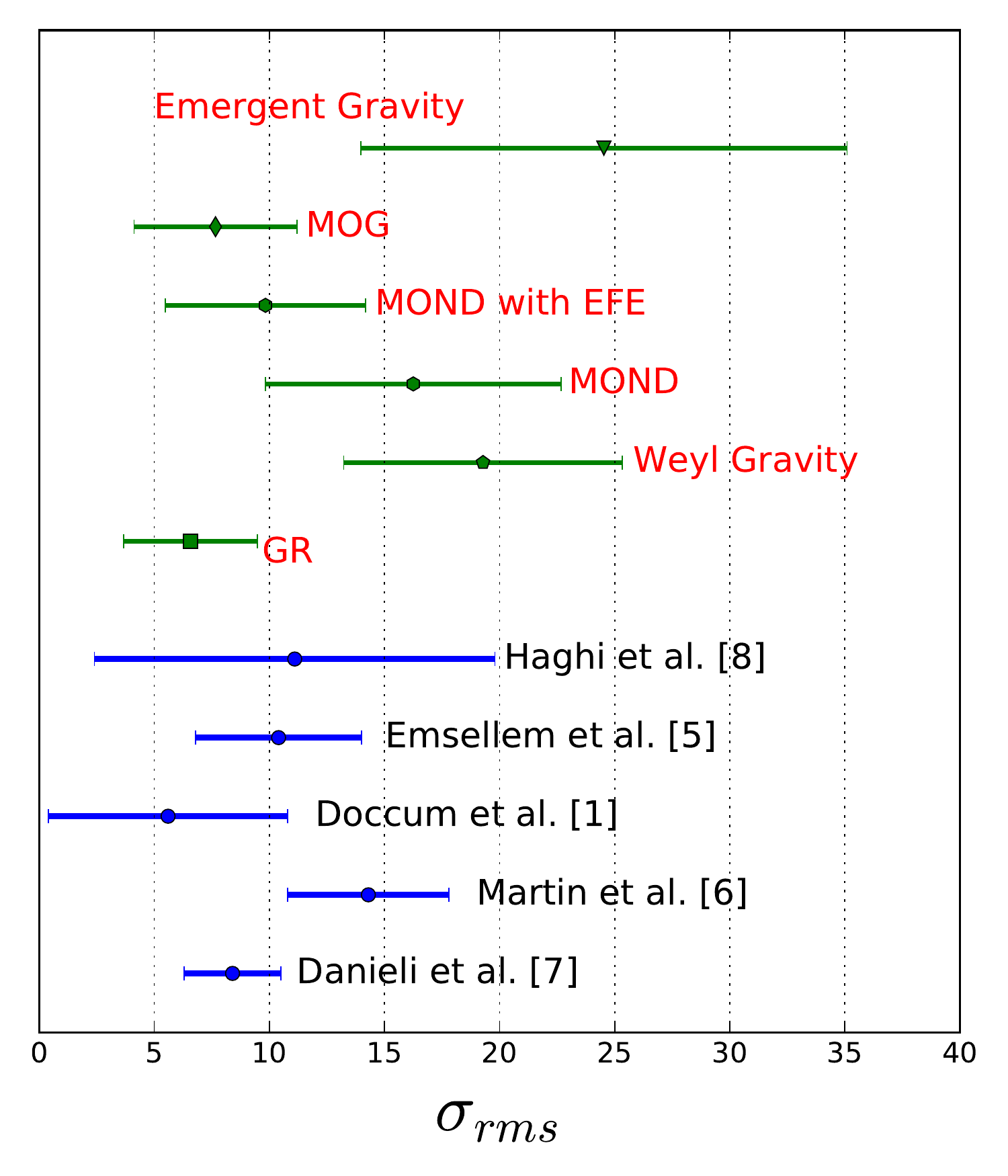}
	\caption[]{\textbf{Estimated bounds on the dispersion profiles}: We report the intrinsic dispersion velocity of  DF2 (in blue) estimated by  vanDokkum \textit{et al.} \cite{van2018galaxy}, Emsellem \textit{et. al} \cite{emsellem2019ultra}, Danieli \textit{et. al} \cite{danieli2019still}, Martin \textit{et. al} \cite{martin2018current} and Haghi \textit{at. al} \cite{haghi2019new}. Different statistical methods have been employed in these studies. Along with those, we show the computed bounds on the rms velocity dispersion for different theories of gravity (in green; this work).}
	\label{fig4}
\end{figure}
In Figure  \ref{fig4}, we show the root mean square (rms) velocity dispersion of the observed profile, estimated by different groups employing different statistical methods \cite{van2018galaxy,emsellem2019ultra,martin2018current,danieli2019still,haghi2019new}; and compare with that of the predicted rms velocity dispersion (from baryonic mass profile only) in different theories of gravity. We compute the rms velocity dispersion generated by the baryonic mass distribution of the galaxy:
\begin{equation}
\sigma_{rms}^2=\frac{\int_0^{R_{cut}} \sigma^2(r')  r'^2\rho_{N}(r')~dr'}{\int_0^{R_{cut}} r'^2\rho_{N}(r')~dr'},
\label{eqnrms}
\end{equation}
where $\rho_{N}$ is the number density. We take $\rho_{N}=\rho / M_{\odot}$. The error-bars on the estimated rms velocity dispersion of the predicted modfiied gravity profiles have been obtained by varying the effective radius by $50\%$. Such practice is very similar in spirit to the case where one computes the rms dispersion using the minimum and maximum values for either the mass density parameters or the parameters in respective theories of gravity and report the resultant bounds.  We find that within the uncertainties of measurements, both GR and modified gravity theories (except Emergent gravity) match with the observation. To be more specific, all of the gravity models in question are in perfect harmony with the most recent inferred intrinsic velocity dispersion by Haghi \textit{at. al} \cite{haghi2019new} at the $2\sigma$ confidence level. Predicted bounds in MOG, MOND with EFE and GR is more consistent with the estimates reported by vanDokkum \textit{et al.} \cite{van2018galaxy}, Emsellem \textit{et. al} \cite{emsellem2019ultra} and Danieli \textit{et. al} \cite{danieli2019still}. Other modified gravity models match these estimates only in the lower limits. On the other hand, both GR and MOG bounds are slightly lower than the values inferred by Martin \textit{et. al} \cite{martin2018current} whereas other modified gravity theories are fully consistent with them. 
\begin{figure}[ht]
	\centering
	\includegraphics[width=1.0\linewidth]{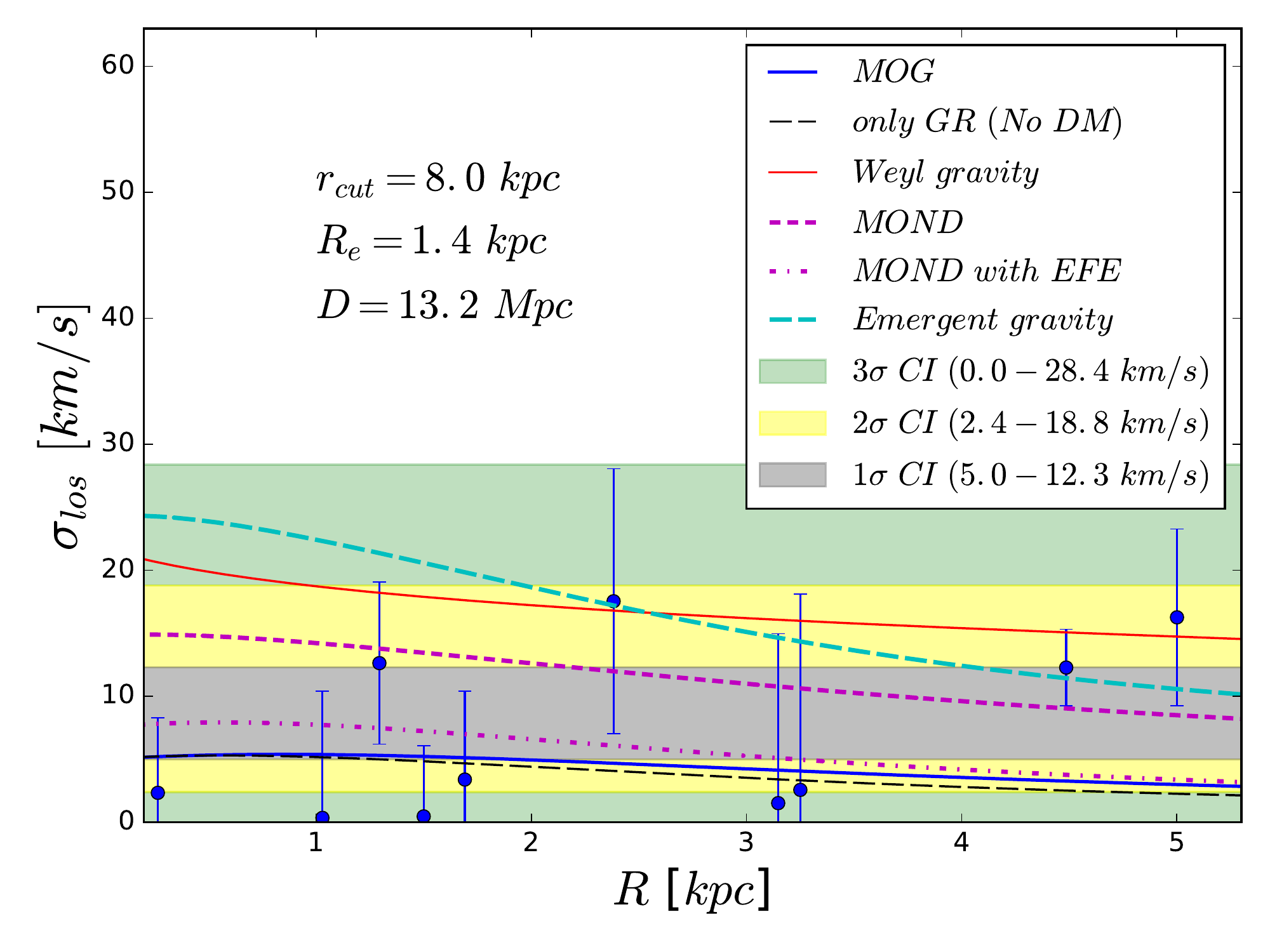}
	\caption[]{\textbf{Line-of-sight velocity dispersion profile of NGC1052-DF2 for $D=13.2$ Mpc}: Details remain same as in Figure \ref{fig2}.}
	\label{fig3}
\end{figure}

\begin{table}[h]
	\centering
	\caption[Milky Way mass model]{\textbf{Reduced chi-square values as goodness-of-fits for different theories of gravity}. $D=13.2$ $Mpc$ and no dark matter are assumed.}
	\label{T2}
	\begin{tabular}{c c}
		\hline 
		\hline
		&$\chi^{2}/dof$ \\
		General Relativity (GR) without dark matter &2.04\\
		Modified Gravitational Theory (MOG) & 1.81\\
		Modified Newtonian Dynamcies (MOND) & 1.89\\
		MOND with EFE & 1.79\\
		Weyl Conformal Gravity & 3.35\\
		Emergent Gravity & 4.55\\
		\hline 
		\hline
	\end{tabular} 
\end{table}

At this point, we ask how our results will change (if at all)  as the assumption (or measurement) of the distance to the galaxy NGC1052-DF2 changes. We therefore perform the same analysis for $D=13.2$ Mpc and report the findings in Figure \ref{fig3}. The galacto-centric distance to the individual GC is rescaled with D (as different D will yield a different conversion factor between the angular separation and projected radial distances). Though we do not specifically know, whether NGC1052-DF2 is expected to have any likely association with any massive galaxy at this distance, we assume a generic trimming radius $r_{cut}=8$ kpc. We find that all modified gravity theories yield a better match at this distance, compared to earlier analysis assuming $D=20$ Mpc. Furthermore, the match between data and Emergent gravity profile becomes comparable to MOND and Weyl gravity. Reduced chi-square values for different theories are shown in Table \ref{T2}.


\section{\textbf{Discussion \& Conclusion}}
\label{sec6} In this paper, we present the \textit{first} analysis of detailed radial variation of the velocity dispersion profile of the ultra-diffuse galaxy NGC1052-DF2 in the context of GR (without dark matter), MOND (for both isolated and EFE cases), MOG, Weyl conformal gravity and Emergent gravity. We show that GR, MOG and MOND with EFE produce excellent match to the observed data while (isolated) MOND and Weyl gravity gives acceptable fits. Emergent gravity, however, performs poorly in fitting the observed dispersion data if distance, $D$, is assumed to be 20 Mpc. All other theories agree with the data within $2\sigma$ ($3\sigma$ for Weyl gravity) confidence level for both the distance measurements ($D=[13.2,20.0]$ Mpc) reported in literature. The closer value of $D=13.2$ Mpc results a better match between modified gravity theories and data.

\begin{figure}[ht]
	\centering
	\includegraphics[width=1.0\linewidth]{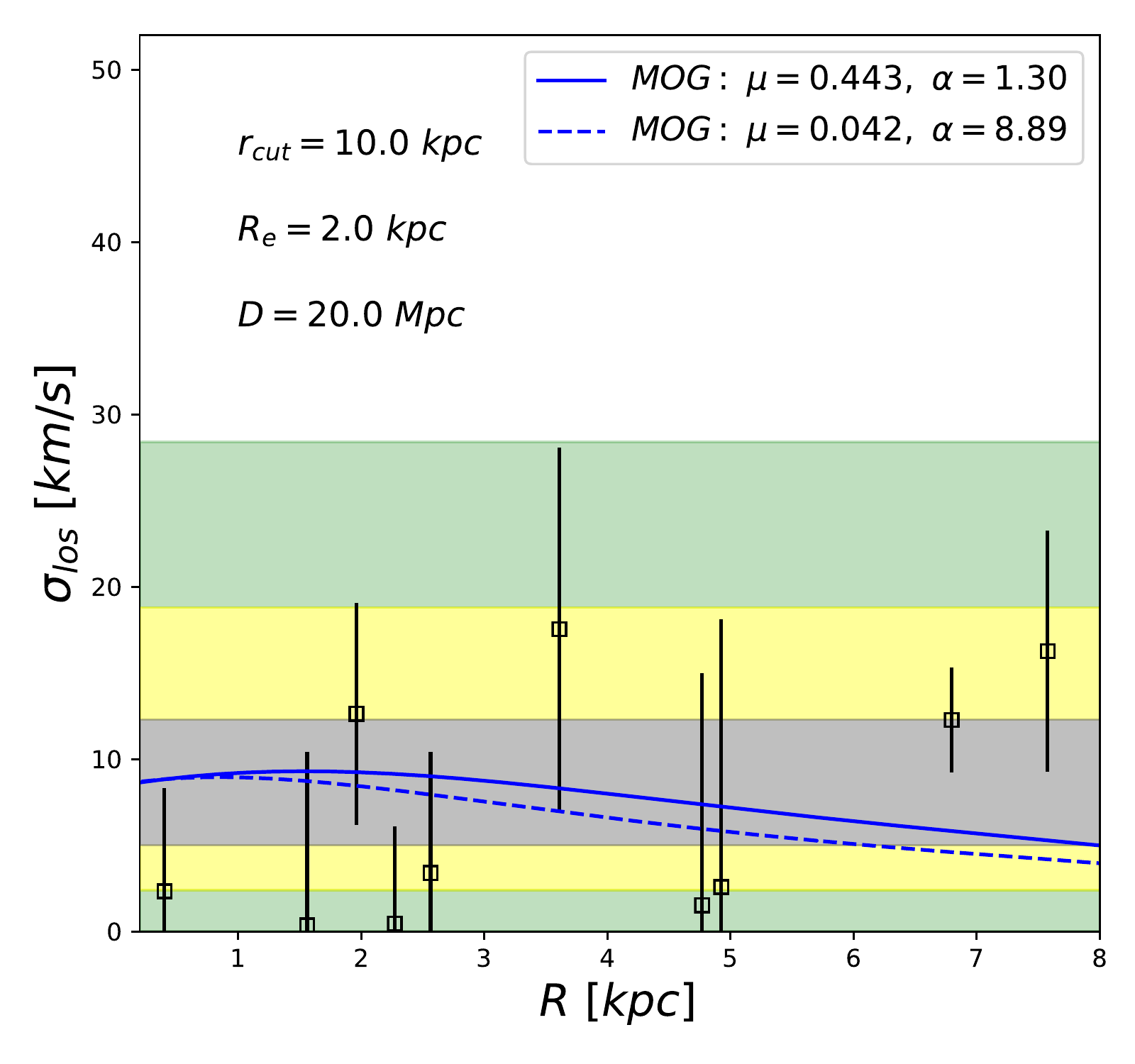}
	\caption[]{\textbf{Line-of-sight velocity dispersion profile of NGC1052-DF2 for $D=20$ Mpc}: We plot the predicted MOG profiles for two different sets of parameter values $\alpha=1.30$ and $\mu=0.443$ $kpc^{-1}$ $kpc^{-1}$ (solid blue) \& $\alpha=8.89$ and $\mu=0.042$ $kpc^{-1}$ (dashed blue). Details remain same as in Figure \ref{fig2}.}
	\label{fig5}
\end{figure}

One interesting aspect of our result is that MOG produces a better match with data \cite{wasserman2018deficit} compared to other modified gravity theories. In fact, predicted dispersion profile in MOG is very similar to that of GR (without dark matter). The reason lies in the fact that, unlike other modified gravity theories we investigate in this paper, MOG enjoys one additional degree of freedom in its structure. While the parameter values in Weyl gravity, MOND and Emergent gravity are same for all galaxies, parameters in MOG depends on the mass and length scale of the system. This allows MOG to fit the data more efficiently than other theories. It must be noted that, in this study, we set the MOG parameter values $\alpha=1.30$ and $\mu=0.443$ $kpc^{-1}$. These values of $\alpha$ and $\mu$ are different than the values obtained through galaxy rotation curve fits: $\alpha=8.89$ and $\mu=0.042$ $kpc^{-1}$. In fact, these two parameters are shown to depend on the system mass through the scaling $\alpha = \alpha_{\inf} \frac{M}{(\sqrt{M}+E)^2}$ and $\mu = \frac{D}{\sqrt{M}}$ where $\alpha_{\inf}=10$, $D=6.25\times 10^3$ $M_{\odot}^{1/2} $ $kpc^{-1}$ and $E=2.5\times 10^4$ $M_{\odot}^{1/2} $ \cite{moffat2009fundamental}. For NGC 1052-DF2, the values for $\alpha$ and $\mu$ turn out to be: $\alpha=1.30$ and $\mu=0.443$ $kpc^{-1}$ \cite{moffat2018ngc}. For the sake of completeness, in Figure \ref{fig5}, we plot the resultant MOG profiles with these two sets of choices for $\alpha$ and $\mu$. We find that both the choices give good fit with the data.

We further investigate whether the choice of cutoff radius $r_{cut}$ has any significant effect on the predicted dispersion profiles; particularly for Weyl gravity. Eq. \ref{weylequ} suggests that the predicted acceleration in Weyl gravity might be sensitive to $r_{cut}$ given there are terms such as $I_2$ involving 4-th moments of the density.  We, however, find out that the mass density itself drops rapidly (exponentially) which ensures even faster drop for the 4-th moments of the density. Therefore, a higher value of $r_{cut}$ will only slightly increase the dispersion values in the outskirt of the galaxy. This will have little effect on our results.

We note that while this work has been is progress, vD19 \cite{van2019second} claimed to find a second galaxy, named NGC1052-DF4, at the same distance $D=20$ Mpc without the `dark matter'. The effective radius of the galaxy is $R_e = 1.0$ kpc and total luminosity is $L=7.7\times10^{8}$ $L_{\odot}$ from which vD19 inferred a total mass of $M\sim1.5\times 10^{8}M_{\odot}$ \cite{van2019second}. We approximately model the galaxy NGC1052-DF4 with  a mass profile similar to NGC1052-DF2 (Equation \ref{mass}) but for $\varSigma_0=1.15 \times 10^7 M_{\odot}/kpc^2$ and $R_e=1.0$ $kpc$. The resultant rms values of the velocity dispersion in different theories are presented in Table \ref{T3}. We find that all of the modified gravity theories agree with the inferred intrinsic velocity dispersion value either at $90\%$ (GR, MOG, MOND and MOND with EFE) while Weyl gravity and Emergent gravity lies a bit outside $95\%$ confidence level if $D=20$ Mpc is assumed. However, we point out that even for this galaxy the distance measurement of vD19 has been refuted by Monelli \& Trujillo \cite{monelli2019trgb} who reported $D=13.5$ Mpc. At this distance, the galaxy exhibit properties similar to other dwarf galaxies \cite{monelli2019trgb} and thus modified gravity theories will easily be able to account for the observed dispersion profiles. A detailed study of the radial variation of dispersion profiles, similar to one presented for NGC1052-DF2 in this paper, would be needed to draw strong inference regarding the ability of modified gravity theories in fitting DF4 dispersion profile. We leave this for a future exploration.

\begin{table}[h]
	\centering
	\caption[]{\textbf{Predicted rms value of the velocity dispersion profiles in different theories of gravity for NGC1052-DF4}. $D=20.0$ $Mpc$ and no dark matter are assumed. The quoted $1\sigma$, $2\sigma$ and $3\sigma$ values for intrinsic velocity dispersion are taken from \cite{haghi2019new} which uses data reported in vD19 \cite{van2019second}.}
	\label{T3}
		\begin{tabular}{c c}
			\hline 
			\hline
			&$\sigma$ (km/s) \\
			\textbf{Inferred from Observation} &\\
			1$\sigma$ confidence level  & $<\sim$7.6\\
			2$\sigma$ confidence level  & $<\sim$13.0\\
			3$\sigma$ confidence level  & $<\sim$17.5\\
			\textbf{Prediction from Theories with $D=20$ Mpc} & \\
			General Relativity (GR) without dark matter &4.91\\
			Modified Gravitational Theory (MOG) & 5.60\\
			Modified Newtonian Dynamcies (MOND) & 11.67\\
			MOND with EFE & 7.32\\
			Weyl Conformal Gravity &13.53\\
			Emergent Gravity & 16.45\\
			\hline 
			\hline
		\end{tabular} 
\end{table}

Our analysis considers the mass profile to be spherical. Any departure from our assumption in the GC/stars profile can slightly alter the predicted dispersion profile. The uncertainties regarding the inclination angle of the galaxy would also result some error in our estimation of dispersion velocities.  On top of that, we have used a generic value for the trimming radius $r_{cut}$ (to model tidal interaction) for all modified theories gravity. The value of $r_{cut}$ could in principle vary in different theories of gravity. These simplifications would unlikely to alter the final conclusion much. However, we note that our analysis assumes the dispersion profile to be isotropic ($\xi=0.0$) i.e. there exists no difference between the dispersion velocities in azimuthal and radial directions. In reality, galaxies can exhibit non-zero anisotropy values. It would be interesting to investigate whether relaxing the assumption of $\xi=0$ helps some modified gravity theories (e.g. Weyl gravity or Emergent gravity) to better fit the data. A detailed systematic investigation of these issues is beyond the scope of this paper and would be carried out in future.

In summary, we demonstrate that NGC1052-DF2 (and NGC1052-DF4) does not falsify modified gravity theories in general. In fact, current observation can be interpreted either as an evidence for a galaxy without (or little) dark matter or as a weird ultra-diffuse galaxy which re-confirms the success of modified gravity paradigm at the galactic scale.

\noindent \textbf{Acknowledgement:}
TI thanks the Long Term Visiting Students Program at ICTS-TIFR where the work began. TI is supported by a Doctoral Fellowship at UMass Dartmouth. TI's research is, in addition, partially supported by NSF grant PHY1806665. KD would like to thank Abdus Salam ICTP, Trieste, Italy for their hospitality during the course of the work. The authors sincerely thank J. W. Moffat, Pavel Kroupa, Keith Horne and Hongsheng Zhao for their comments on the manuscript; Eric Emsellem for correcting mistakes in the quoted dispersion values; and the anonymous referees for their suggestions towards improving the MOND analysis. 

\end{document}